\documentstyle[11pt,moriond,epsfig,amssymb,cite]{article}

\def\ifmath#1{\relax\ifmmode #1\else $#1$\fi}%
\def\rd{\ifmath{{\mathrm{d}}}}
\def\re{\ifmath{{\mathrm{e}}}}

\def\rg{\ifmath{{\mathrm{g}}}}

\def\rr{\ifmath{{\mathrm{r}}}}
\def\rR{\ifmath{{\mathrm{R}}}}
\def\rI{\ifmath{{\mathrm{I}}}}
\def\rrm{\ifmath{{\mathrm{m}}}}
\def\rT{\ifmath{{\mathrm{T}}}}
\def\rW{\ifmath{{\mathrm{W}}}}
\def\rq{\ifmath{{\mathrm{q}}}}
\def\BE{\ifmath{{\mathrm{BE}}}}
\def\ch{\ifmath{{\mathrm{ch}}}}
\def\incl{\ifmath{{\mathrm{incl}}}}
\def\out{\ifmath{{\mathrm{out}}}}
\def\output{\ifmath{{\mathrm{output}}}}
\def\input{\ifmath{{\mathrm{input}}}}
\def\side{\ifmath{{\mathrm{side}}}}
\def\rL{\ifmath{{\mathrm{L}}}}

\def\be{\begin{equation}}
\def\ee{\end{equation}}
\def\bea{\begin{eqnarray}}
\def\eea{\end{eqnarray}}

\def\G{\Gamma}
\def\d{\delta}
\def\D{\Delta}

\def\k{\kappa}
\def\la{\lambda}

\def\r{\rho}

\def\S{\Sigma}
\def\Y{\Psi}

\def\cent{\centerline}

\def\hs{\hskip}
\def\vs{\vskip}
\def\ni{\noindent}

\def\ran{\rangle}
\def\lan{\langle}

\begin{document}
\hfill HEN-420
\vspace*{4cm}
\title{INTERCONNECTION EFFECTS AND W$^+$W$^-$ DECAYS\\
 (a critical (p)(re)view)\ \footnote{Invited talk at the XXXIVth
Rencontre de Moriond, QCD and High Energy Hadronic Interactions,
Les Arcs (France), March 20-27, 1999 (full version)}}

\author{ W. KITTEL}

\address{HEFIN, University of Nijmegen/NIKHEF, \\
Toernooiveld 1, 6525 ED Nijmegen, The Netherlands}

\maketitle\abstracts{
Color reconnection and Bose-Einstein correlations not only can have
an influence on the measurement of the W-mass in the fully hadronic
W$^+$W$^-$ decay channel at LEP2, but also can give essential information
on the structure of the QCD vacuum and the space-time development of a 
$q_1\bar q_2$ system.
Recent developments are critically analyzed, with particular emphasis
on the models used in this field. More sensitive variables are needed to
distinguish between color reconnection models, while more experimental
knowledge has to be built into the Bose-Einstein models and, above all,
these two closely related phenomena have to be treated in common. Both 
effects are determined by the space-time overlap of the W$^+$ and W$^-$
decay products. Vital experimental information on the space-time development
of the decay of the $q_1\bar q_2$ system is becoming available from
the high-statistics data on hadronic Z decay and models will have to be
able to explain this evidence before being used to predict interference
effects in hadronic W$^+$W$^-$ decay.}

\section{Introduction}

When Bo Andersson reported on the incorporation of Bose-Einstein correlations
into the Lund string during an earlier workshop,\cite{1} he started:
``this is the most difficult work I have ever participated in'' and
he did not even refer to the W$^+$W$^-$ overlap! The statement sets the
scale, but should be squared when applied to the latter. That's why it is
easier to be {\it critical} on this topic than to {\it review} it and
why I shall reduce my task at this Rencontre to giving a personal (though 
still critical) {\it view}, instead.

Interconnection effects, at first sight a nuisance when trying to measure
the W mass, on the other hand may open new handles for the study of basic
issues as the structure of the vacuum and the space-time development of
a $\rq\bar \rq$ system at high energy.

Of course, the phenomenon of color reconnection is by no means restricted to 
W$^+$W$^-$ decay. Other examples are J/$\Y$ production in B decay, 
event shapes in Z-decay or rapidity gaps at HERA.

\section{Color reconnection}
\subsection{The models}

If produced in the same space-time point, pairs of quarks and anti-quarks 
($\rq_1\bar \rq_4$) and $(\rq_3\bar \rq_2)$ originating from the decay 
of {\em different} W's can form strings, if they happen to be in a color 
singlet.\cite{2} From color
counting, this is fulfilled in $1/9$ of the cases, but this recoupling
probability can be enhanced by gluon exchange.
However, the pairs $(\rq_1\bar \rq_2)$ and $(\rq_3\bar \rq_4)$ 
are produced at a distance $\propto 1/\G_\rW\approx 0.1$ fm, small 
compared to the hadronic scale, but large enough to suppress exchange 
and/or interference of hard $(E_\rg\gtrsim\G_\rW$) gluon.\cite{3,4}
Soft-gluon interference, is of course possible. It depends on the vacuum
structure and a number of models exist.\cite{3,4,5,6,7,8} According to
the underlying software package used, they can be grouped into the
following.

\vs 2mm
\ni
1. {\em PYTHIA based models:}\\
a) SKI \cite{3} uses Lund strings and allows at most one reconnection.
The color field is treated as a Gaussian-profile flux tube (as in a type I 
superconductor) with a radius of $\sim$0.5 fm and the recoupling probability 
$\r$ depends on the overlap of the two flux tubes in space-time. The
recoupling probability density is a free parameter quite arbitrarily chosen
to be 0.9 fm$^{-4}$ (but varied easily). At 183 GeV, recoupling is predicted
to occur in 38\% of the events.\cite{9}

b) SKII \cite{3} also uses Lund strings with at most one reconnection,
but the color field is treated as an exponential-profile vortex line
(as in a type II superconductor). When two vortex lines cross, i.e. have
a space-time point in common, for the first time, they recouple with unit 
probability $(\r=1)$. At 183 GeV, this gives a recoupling in 22\% of the 
events.\cite{9}

c) SKII' \cite{3}: like SKII' but with $\r=1$ upon first crossing 
reducing the total string length, giving a recoupling in 20\% of the
183 GeV/$c$ events.\cite{9}

d) \v{S}TN \cite{5}: is an important extension of SKI and SKII to implement
the space-time evolution of the shower, as well as multiple reconnection,
including self-interaction of strings.
This approach is more realistic than the SK versions, but still shares one
problem: the color reconnection is performed after the generation of the
complete parton shower and, therefore, cannot change its development.

\vs 2mm
\ni
2. {\em Color-dipole based models} \cite{4,6}:\\
Here, the Lund string and its gluon kinks are replaced by a chain of dipoles.
Within, or between two dipole chains, reconnection is possible when the
color indices (ranging from 1 to 9) of two (non-adjacent) dipoles are the
same. Reconnection is indeed performed when the string length measure
$\la=\sum^{n-1}_1 \ln (m^2_{i,i+1}/m^2_0)$ is reduced $(m_{i,i+1}$ is the 
invariant mass of the string segment stretched by partners $i$ and $i+1$ and
$m_0$ a hadronic mass scale around 1 GeV).
Also these models exist in a number of versions. In \cite{4}, the number
of reconnections per event was at most one, and there was no 
reconnection within a W. In version \cite{6}, two dipole systems 
$\rq_1\bar \rq_2$ and $\rq_3\bar \rq_4$ first evolve separately, radiating 
gluons with $E_\rg>\G_\rW$ independently, but with color reconnections
within each dipole system. Then, when $E_\rg<\G_\rW$, reconnections between
the two systems are switched on. In practice, because of the
$k_\rT$-ordering of CDM, the cascade is run twice: first for $E_\rg> \G_\rW$ 
without reconnections between the two systems, and a second time allowing
only $E_\rg<\G_\rW$ with interconnections.
An an alternative, the second cascade can be omitted, but interconnections
between the systems is allowed before fragmentation.

\vs 2mm
\ni
3. {\em Cluster models}:
Quarks and gluons originating from the parton showers combine into clusters. 
These are less extended and less massive than strings are and decay 
isotropically into a small number of hadrons.

a) HERWIG based \cite{7}: After showering, the gluons are split 
non-perturbatively into quark-antiquark pairs and each may form a 
color-singlet cluster with a color-connected partner. At the start of the
cluster-formation phase, color connections are established between clusters
that reduce the space-time extension of the clusters, and reconnections are
allowed in $1/9$ of the cases. Reconnection among the products of a
single shower are natural in this model.

b) VNI based \cite{8}: Three scenarios are considered for cluster formation, 
one of which including non-singlet clustering, where the net color of the 
cluster is carried off by a secondary parton.

{\em 
Two critical comments on all models: they should contain reconnection within 
a single W, and if they do, they should be very carefully retuned on the Z.}
Interesting in this connection is an OPAL study \cite{abbi} of gluon production
in Z decay, e$^+$e$^-\to \rq\bar\rq \rg_\incl$, where reconnection effects
are expected to contribute \cite{4}.
Two versions of the dipole model tested predict noticeably fewer
particles at small rapidities and energies than are observed in the
data (or the conventional QCD programs), as well as a downward shift of
about one unit in the $\rg_\incl$ charged-particle multiplicity.

\subsection{The data}

The recent data are beautifully summarized by the previous speaker,\cite{10}
so that I can restrict myself to a few comments.

Fig.~1 reproduces a comparison of OPAL data \cite{9} and model predictions for
the charged-particle multiplicity (a) and (b) and thrust distributions (c).
The full lines correspond to model versions without reconnection, the other
lines to models with reconnection. Two conclusions from this figure are:

\ni
1. The VNI based model \cite{8} is way off, it does not fit the data at all,
but the simulations are also in strong disagreement with results published in
\cite{8}, which, in their turn, were equally far off (at least in thrust $T$),
but in the opposite direction. Furthermore, the MC code is reported not
to conserve energy.\cite{9} I leave it to the reader to decide to do 
something about this or to forget the model.

\begin{figure}
\begin{minipage}[b]{6.5cm}
\epsfig{file=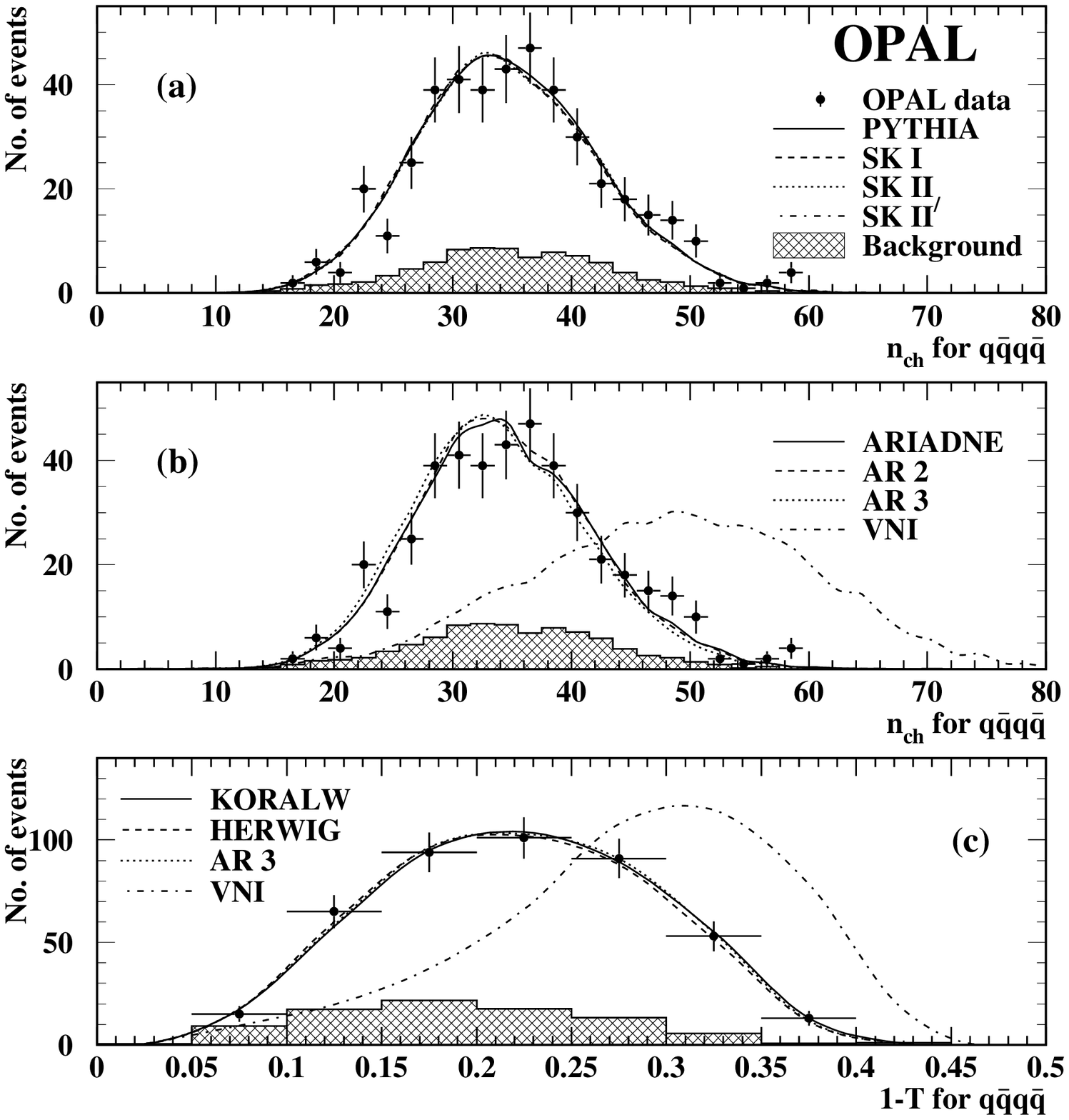,width=8truecm}
\end{minipage}
\hskip1cm
\begin{minipage}[b]{7.5cm}
\epsfig{file=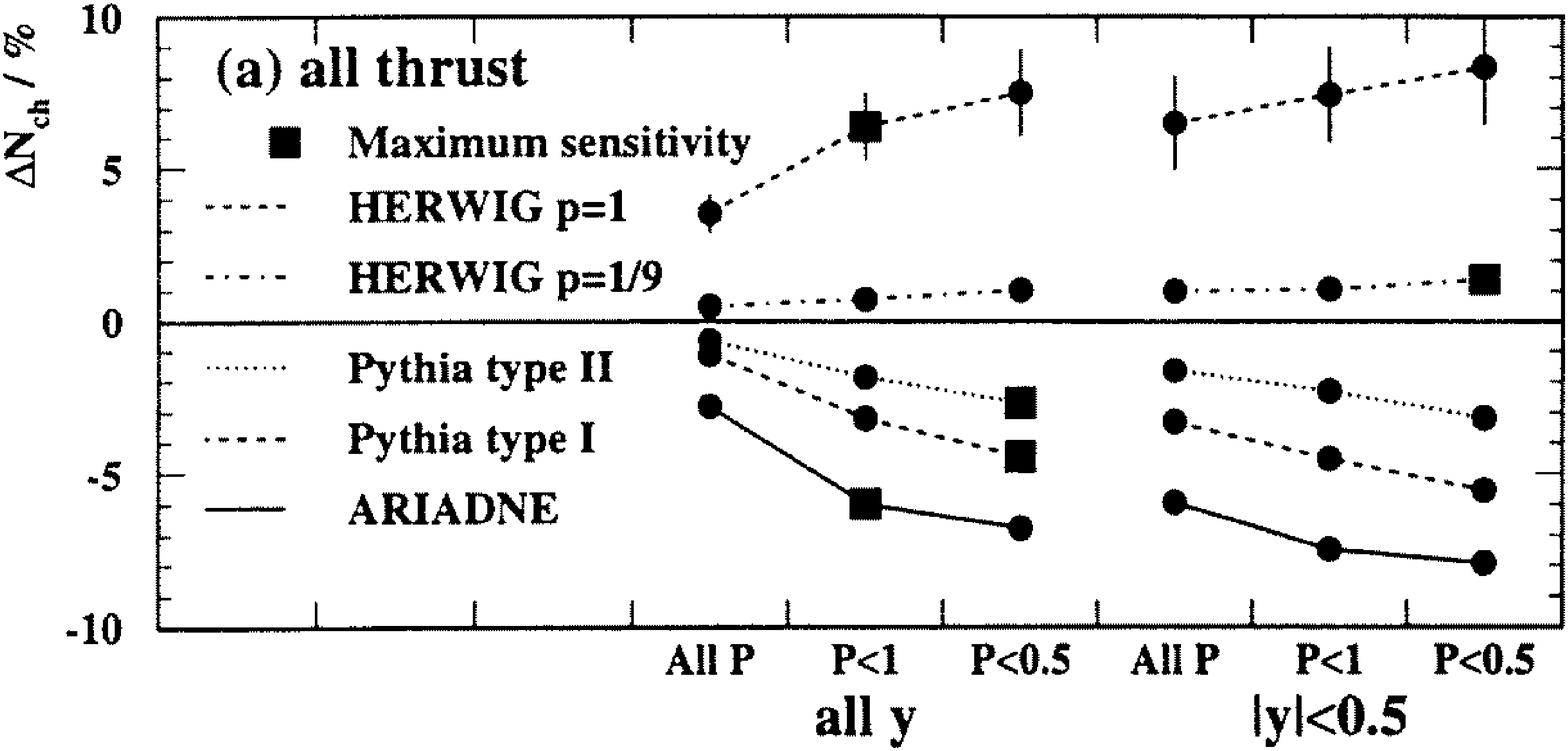,width=8cm}
\vs 1mm
{\footnotesize\baselineskip=10pt\ni
Figure 2: Fractional change in charged-particle multiplicity as a function
of maximum particle momentum, in two rapidity regions.\protect\cite{11}\par}
\vs 8mm
{\footnotesize\baselineskip=10pt\ni
Figure 1: Uncorrected charged-particle multiplicity distribution (a), (b) and
 thrust distribution (c) for e$^+$e$^-\to \rW^+\rW^-\to\rq_1\bar\rq_2\rq_3
\bar \rq_4$ events compared with predictions of models.\protect\cite{9}\par}
\end{minipage}
\end{figure}

\ni 
2. All the other models (including reconnection or not) look so similar
in $n_{\ch}$ and $1-T$, that these variables are obviously not discriminative.

So, the search for color reconnection boils down to the search for 
discriminative variables. As reconnections reduce the string length or the 
cluster size, and these determine the average multiplicity, $\lan n_\ch\ran$
was suspected to be a good candidate. Fig.~1 does, however, not give a
lot of hope, but one can look at $\lan n_\ch\ran$ in the overlap regions,
alone.

In Fig.~2 we reproduce a study of a recent working group.\cite{11} The model
predictions for the multiplicity shift
\be
\D n_\ch = n^{\rW\rW}_\ch - 2 n^\rW_\ch
\ee
are given for all momenta (leftmost points), as well as for a number of
rapidity $y$ and momentum $P$ cuts reducing the sample to that of the
overlap region. The present LEP average for the leftmost point (all $P$,
all $y$) is $\D n_\ch=0.18\pm 0.39$ \cite{10} (or $0.54\pm1.08$\% on Fig.~2).
That means no effect outside errors, but also agreement with color 
reconnection as predicted by PYTHIA and HERWIG.

As the reduced available string length or cluster size will be felt first
by heavy particles, it has been suggested to look at kaons + protons 
with momenta restricted to $0.2-1.2$ GeV/$c$. DELPHI \cite{DELPHI} 
finds a shift of $(+3\pm15)$\%
while ($-8$ to $-3$)\% is predicted.

{\em However, we have to do with a complex overlap of two complex systems,
where correlations and not averages are at play!
Besides that, where the multiplicity is reduced by reducing the string 
length, it is quite likely to be increased by Bose-Einstein correlations! 
The least I recommend, if one wants to restrict oneself to averages, is
to study the shift in integrated {\em two-particle} density, i.e., 
the second-order factorial moment
\be
\D F_2 = F^{\rW\rW}_2-2 F^\rW_2-2\lan n^\rW\ran^2
\ee
or, better, to look at the shift $\D\r(1,2)$ in the two-particle
density itself}.\cite{12}

\section{Bose-Einstein correlations}
The previous speaker \cite{10} has given a beautiful summary on the 
contradictory results on inter-W BE effects. Obviously, before embarking 
on a study of inter-W BE correlation effect, we first have to understand 
intra-W BE correlations and the space-time shape of a {\em single} W. Since 
even that is impossible with present statistics, we have to go back and 
look at the Z$^0$ in more detail!

\subsection{Experimental results on the $Z^0$}

From BE analysis of the Z$^0$ \cite{13} we know, first of all, that BE
correlations indeed exist in its hadronic decay. So, they can, in principle,
give problems in WW overlap. However, more importantly, these very BE 
correlations can be used as a pion-interferometry laboratory to measure the 
space-time development of hadronic Z-decay, and, ultimately in WW overlap. 
It is this, where actually much more is known already than generally used 
in WW studies:

\vs 2mm
\ni
1. {\em Elongation of the pion source} \protect\cite{14,15}:
Applying a two- or three-dimensional (instead of the usual one-dimensional) 
parametrization of the correlation function \cite{16}
\be
R_2(Q_\rL,Q_\out,Q_\side)= 1+\la \exp (-r^2_\rL Q^2_\rL-r^2_\out 
Q^2_\out-r^2_\side Q^2_\side)
\ee
in the longitudinal, out and side components of the squared two-particle
four-momentum difference $Q=(-(p_1-p_2)^2)^{1/2}$ and the corresponding 
size parameters $r_\rL,r_\out,r_\side$, DELPHI \cite{14} and L3 \cite{15} 
find a clear elongation along the thrust axis of the pion source in 
the longitudinal cms.\cite{17} The ratios of the transverse radii ($r_\out,
r_\side)$ and $r_\rT=(r^2_\out + r^2_\side)$ are given in Table 1.
\begin{table}[t]
\caption{Transverse over longitudinal size parameters measured and
predicted by JETSET}
\vs 1mm
\begin{center}
\begin{tabular}{l|cc|c}
         & L3\cite{15} & JETSET & DELPHI (2-jet) \cite{14} \\
\hline
$r_\rT/r_\rL$ & $0.73\pm 0.02^{+0.03}_{-0.10}$ & $0.92\pm0.02$ &
$0.64\pm0.02\pm0.04$ \\
CL(\%)   & 0.8   & $0.4\cdot 10^{-3}$ &   \\
\hline
$r_\out/r_\rL$  & $0.71\pm 0.02^{+0.05}_{-0.08}$ & $0.82\pm0.02$ &
$0.73\pm0.01\pm0.05$ \\
$r_\side/r_\rL$  & $0.80\pm 0.02^{+0.03}_{-0.18}$ & $1.06\pm0.02$ &
$0.58\pm0.01\pm0.02$ \\
CL(\%)   & 3.1   & 0.016 &   \\
\end{tabular}
\end{center}
\end{table}
The elongation is of course stronger in the 2-jet sample used by DELPHI,
but even clear in the full data used by L3. This is in contradiction with the
assumption of a spherically symmetric correlation function in most of the 
models (see below).

\vs 2mm
\ni
2. {\em Position-momentum correlation}:
The correlation length of $0.7<r_\rL<0.8$ fm \cite{14,15} (called the 
{\em length of homogeneity}) corresponds to the length 
in space from which pions are emitted that have momenta similar enough 
to be able to interfere. The spatial extension of the Z emission function 
$S(t,z)$, on the other hand, is expected to be of the order 100 times that 
of $r_\rL$ (see Fig.~3a \cite{18}). 
This invokes a strong momentum ordering in space. 
Experimentally, this emission has been measured only in hadron-hadron 
\cite{19} and heavy-ion collisions,\cite{20} so far (see Fig.~3b,c), 
but its measurement in Z decay will tell us the actual shape of such a decay 
in space-time, and therefore how the overlap of WW has to be visualized! 

\begin{figure}
\hs 3cm a)
\vs -10mm ~
\begin{center}
\epsfig{file=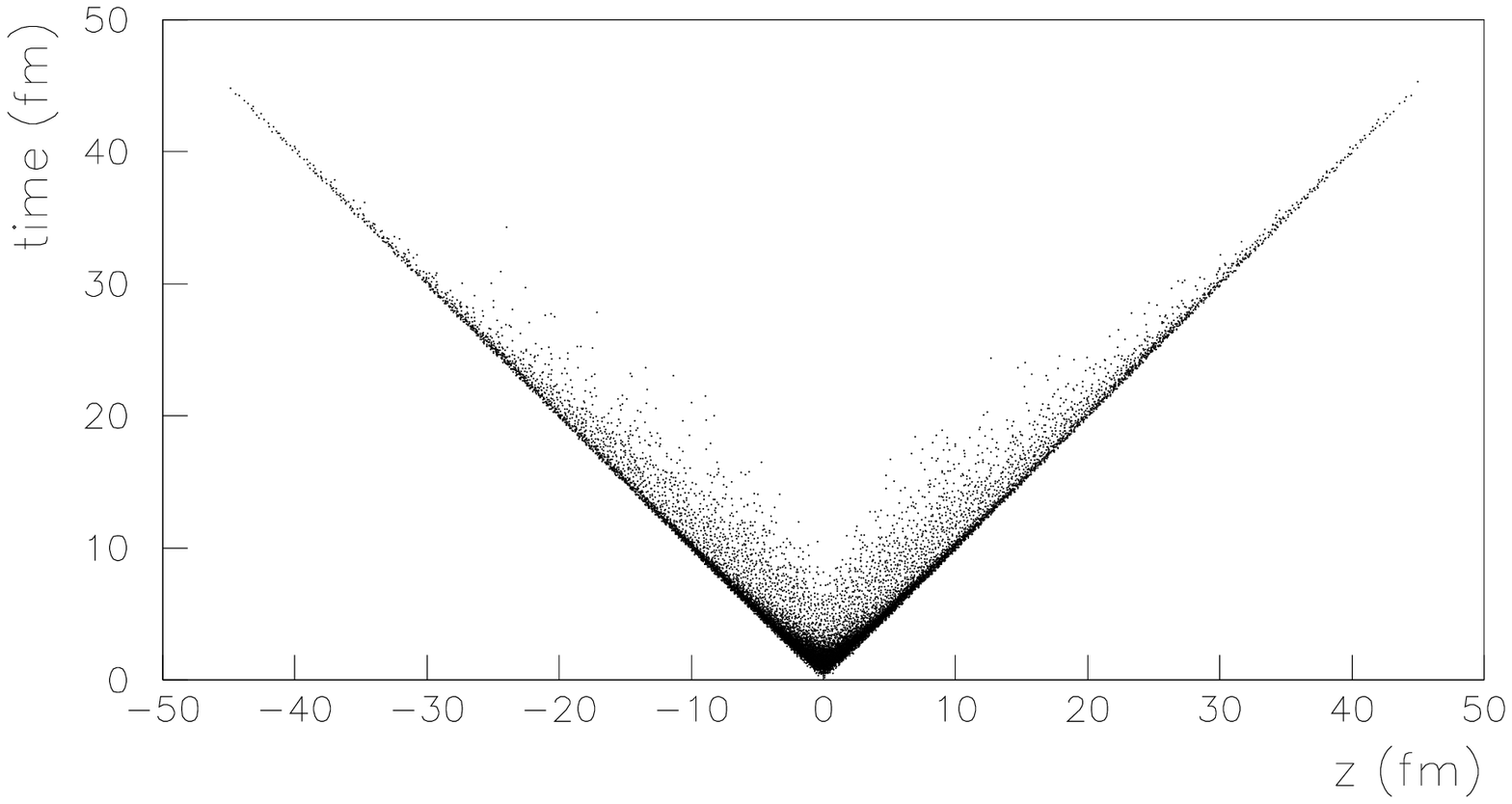,width=7.5cm}
\end{center}
\vs 2mm ~
\begin{minipage}[t]{7.cm}
\leftline {b)}
\vs -5mm ~
\centering\epsfig{figure=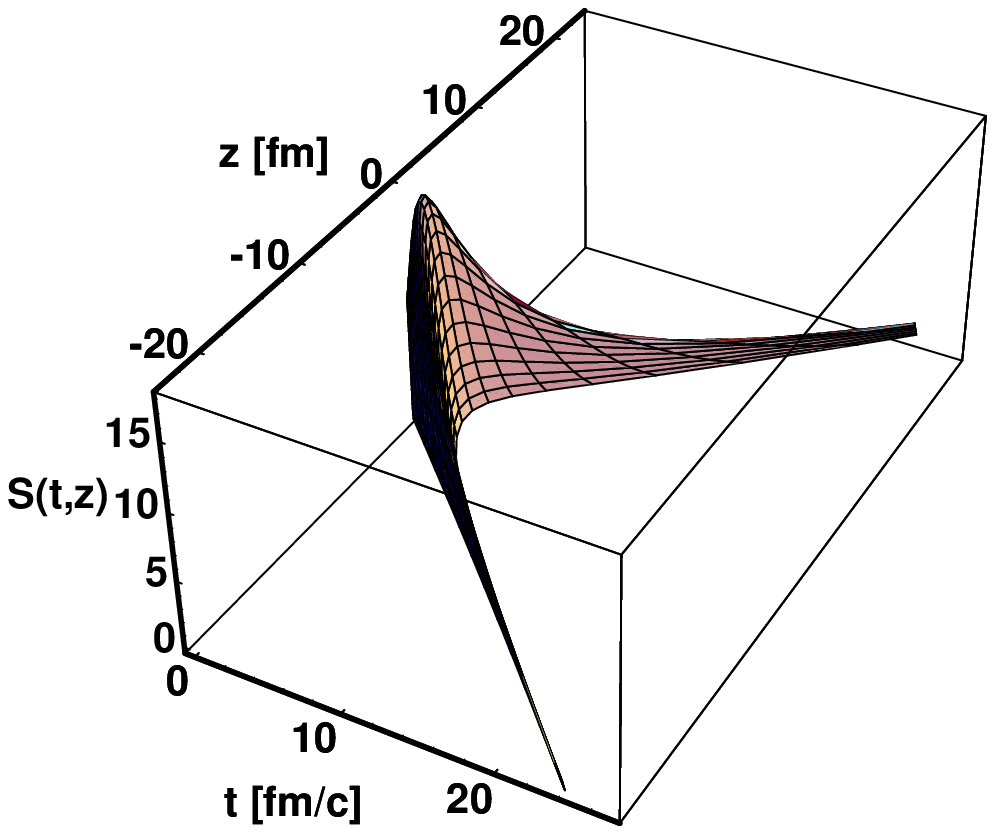,width=7.cm}
\end{minipage}
\hs 4mm
\begin{minipage}[t]{7.2cm}
\leftline{c)}
\centering\epsfig{figure=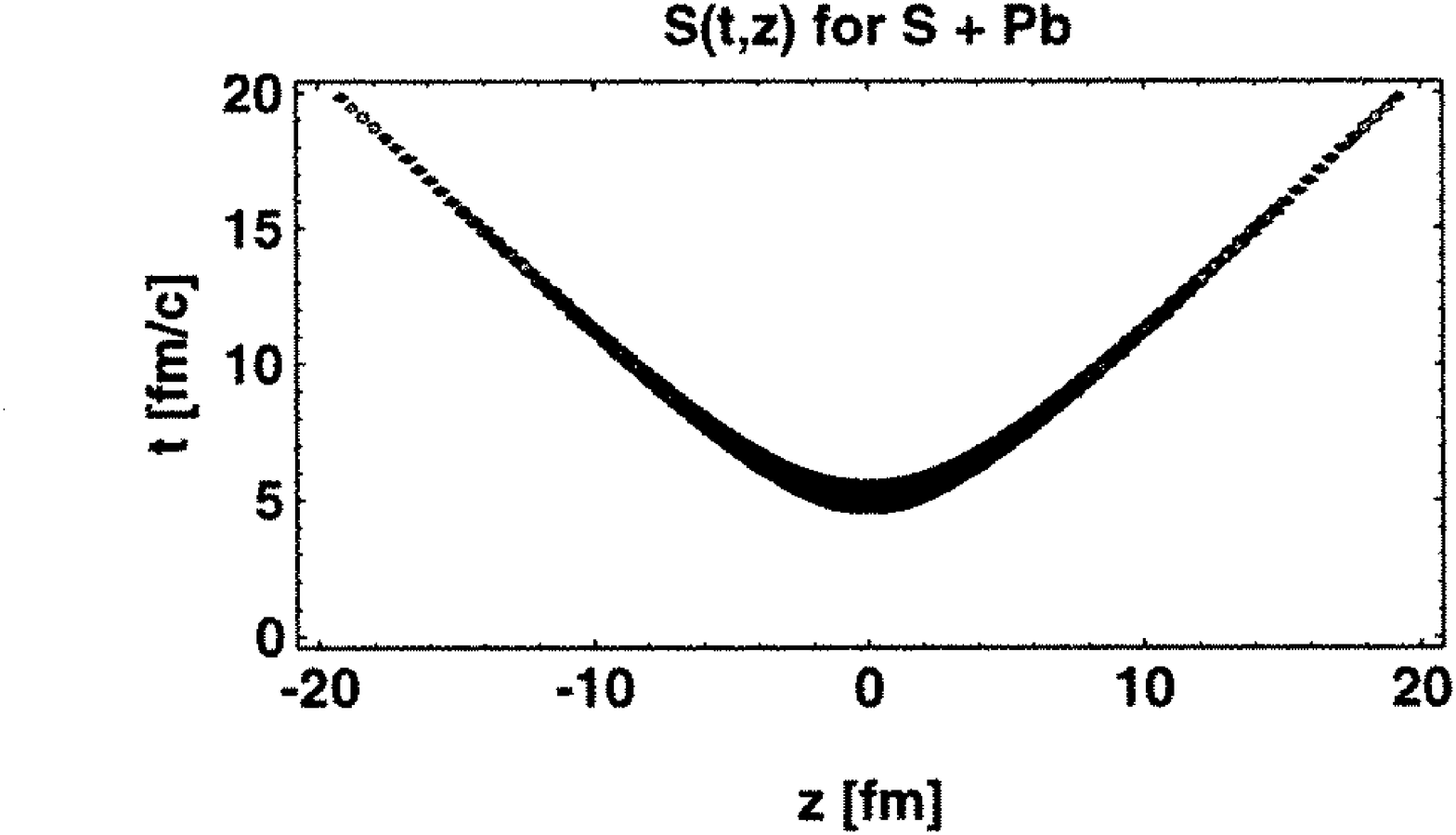,width=9cm}
\end{minipage}
\vs 5mm
  {\footnotesize\baselineskip=10pt\ni
Figure 3.
The emission function $S(t,z)$ as a
function of time $t$ and longitudinal coordinate $z$ for
a) MC for Z$^0$ \protect\cite{18}, b) NA22: \protect\cite{19},
c) NA44 \protect\cite{20}.
\par}
\end{figure}

\vs 2mm
\ni
3. {\em Non-Gaussian correlation function}:
For simplicity, parametrization Eq.~(3), even if not spherically 
symmetric anymore, is still Gaussian. Strong deviation from such
a behavior is known from hadron-hadron collisions,\cite{21} but
deviations also exist at the Z.\cite{13} Generalizing the Gaussian 
to a so-called Edgeworth expansion \cite{22}

\be
R_2=1+\la\prod_i \exp (-r^2_i Q^2_i)\left[ 1+\frac{\k_3}{3!} 
H_3 (r_iQ_i)+\dots\right]\ ,
\ee
where $\k_3$ is the third-order cumulant moment and $H_3$ is the third-order 
Hermite polynomial, shows \cite{15} that the correlation is indeed stronger 
than Gaussian at small $Q$ (see Fig.~4). While maintaining the elongation, 
the CL value of the 3-D fit is now increased from 3.1\% (see Table~1) to 30\%.

\begin{figure}
\cent{\epsfig{file=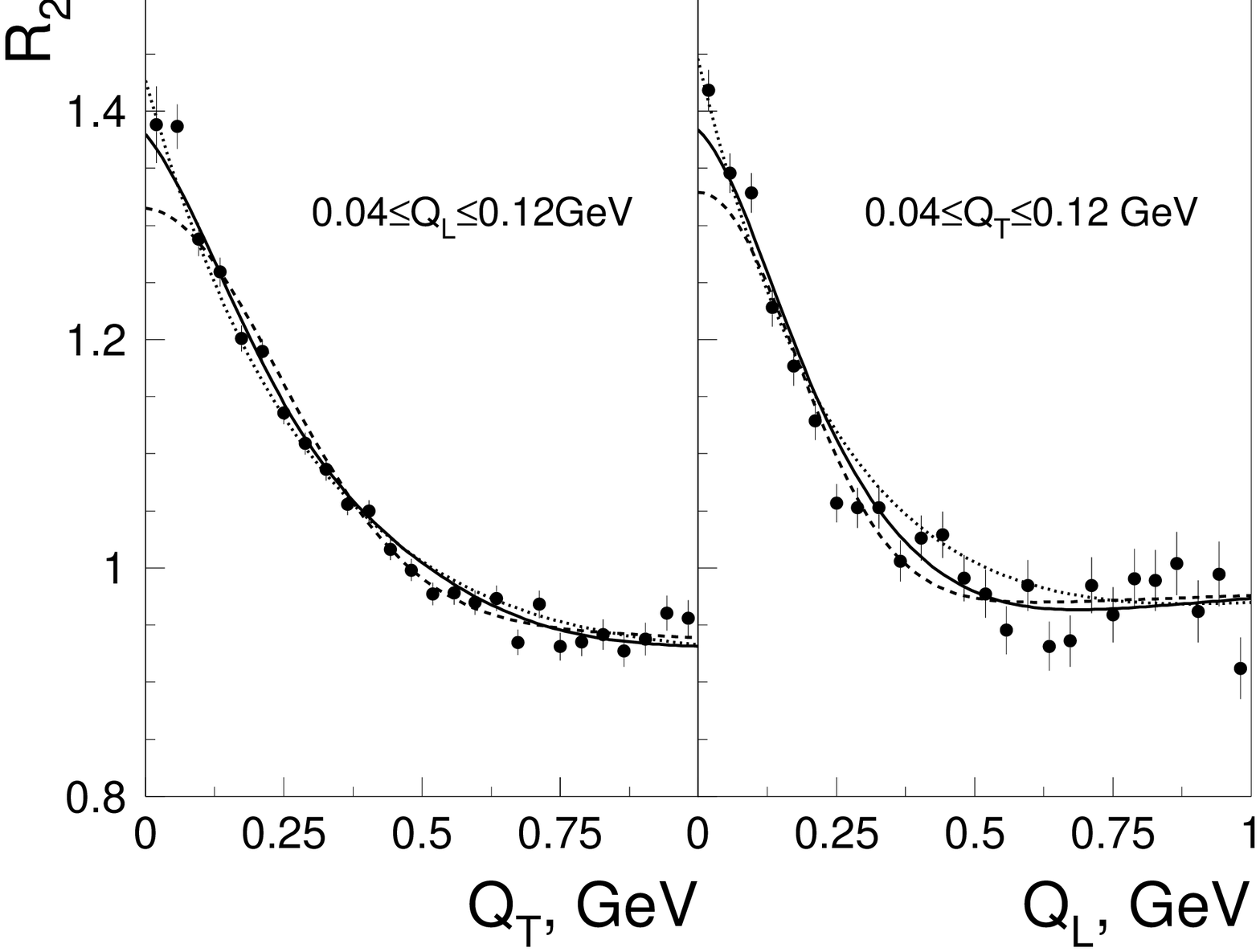,width=10cm}}
\vs 2mm
{\footnotesize\baselineskip=10pt\ni
Figure 4. The BE correlation function $\rR_2$ for hadronic Z decay as a 
function of $Q_\rT=(Q^2_\side+Q^2_\out)^{1/2}$ and $Q_\rL$ after the 
indicated cut in the other variable, compared to fits
by a Gaussian (dashed), exponential (dotted) and Edgeworth expansion
(full).\protect\cite{15}
\par}
\end{figure}

\vs 2mm\ni
4. {\em The transverse mass dependence}:
From heavy-ion collisions,\cite{23} it is known that the radii in Eq.~(3)
decrease with increasing average transverse mass $m_\rT$ of the particle
pair. Preliminary results \cite{14,15} indicate that such a behavior is also
present in Z decay. The $m_\rT$ dependence is reproduced by JETSET/LUBOEI
for $r_\out$, but not for the two other components (see Fig.~5).

\begin{figure}
\cent{\epsfig{file=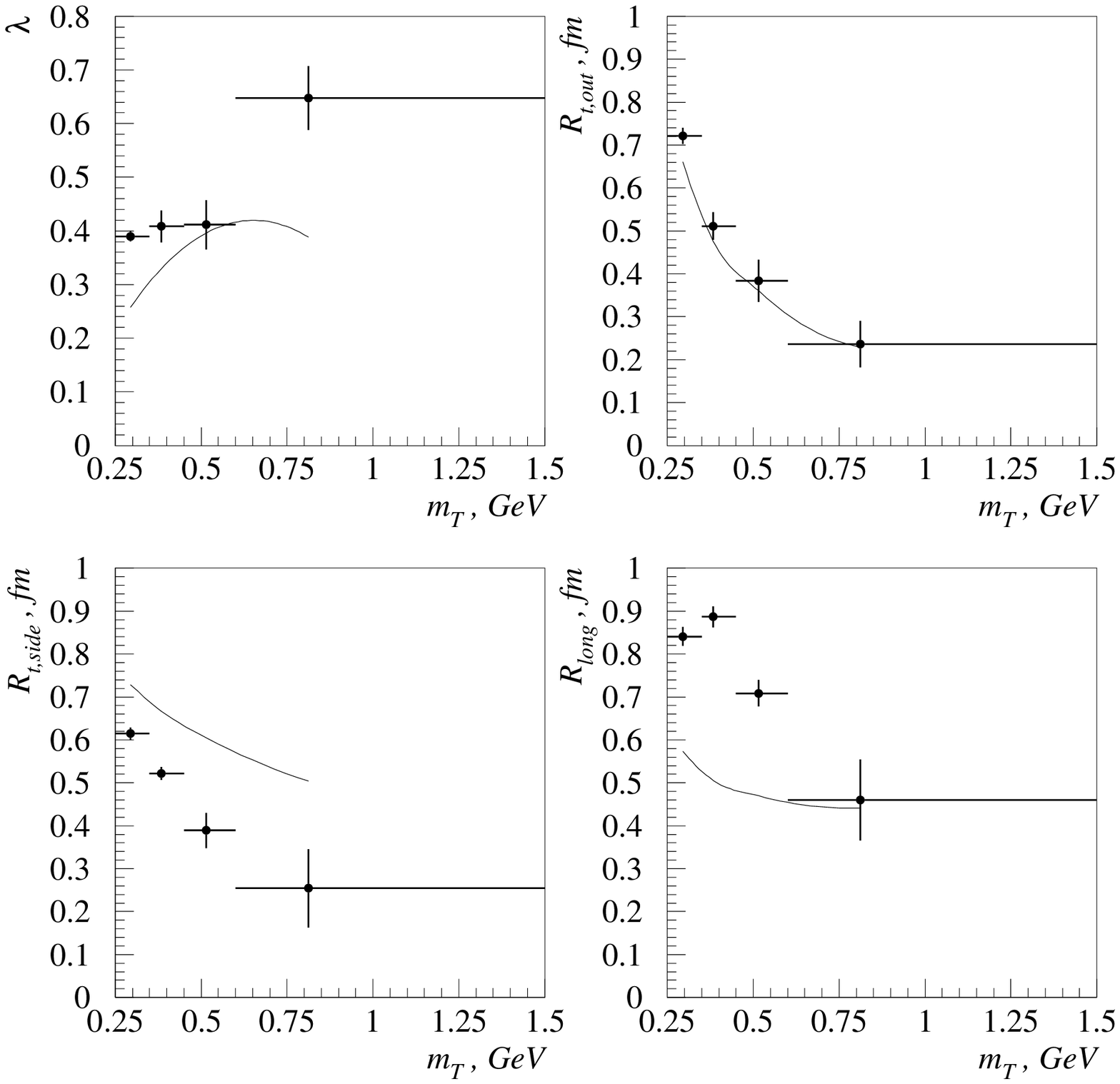,width=10cm}}
\vs 2mm
{\footnotesize\baselineskip=10pt\ni
Figure 5. Transverse mass $m_\rT$ dependence of the three-dimensional
correlation function for hadronic Z decay compared to JETSET 
(full line).\protect\cite{14}\par}
\end{figure}

\vs 2mm\ni
5. {\em Genuine higher-order correlations}
exist in hadron-hadron collisions \cite{21,24} and also in Z 
decay.\cite{25,26} The DELPHI results are given in Fig.~6. For identical 
particles, they are not reproduced by JETSET/LUBOEI.\cite{25}

\begin{figure}
\cent{\epsfig{file=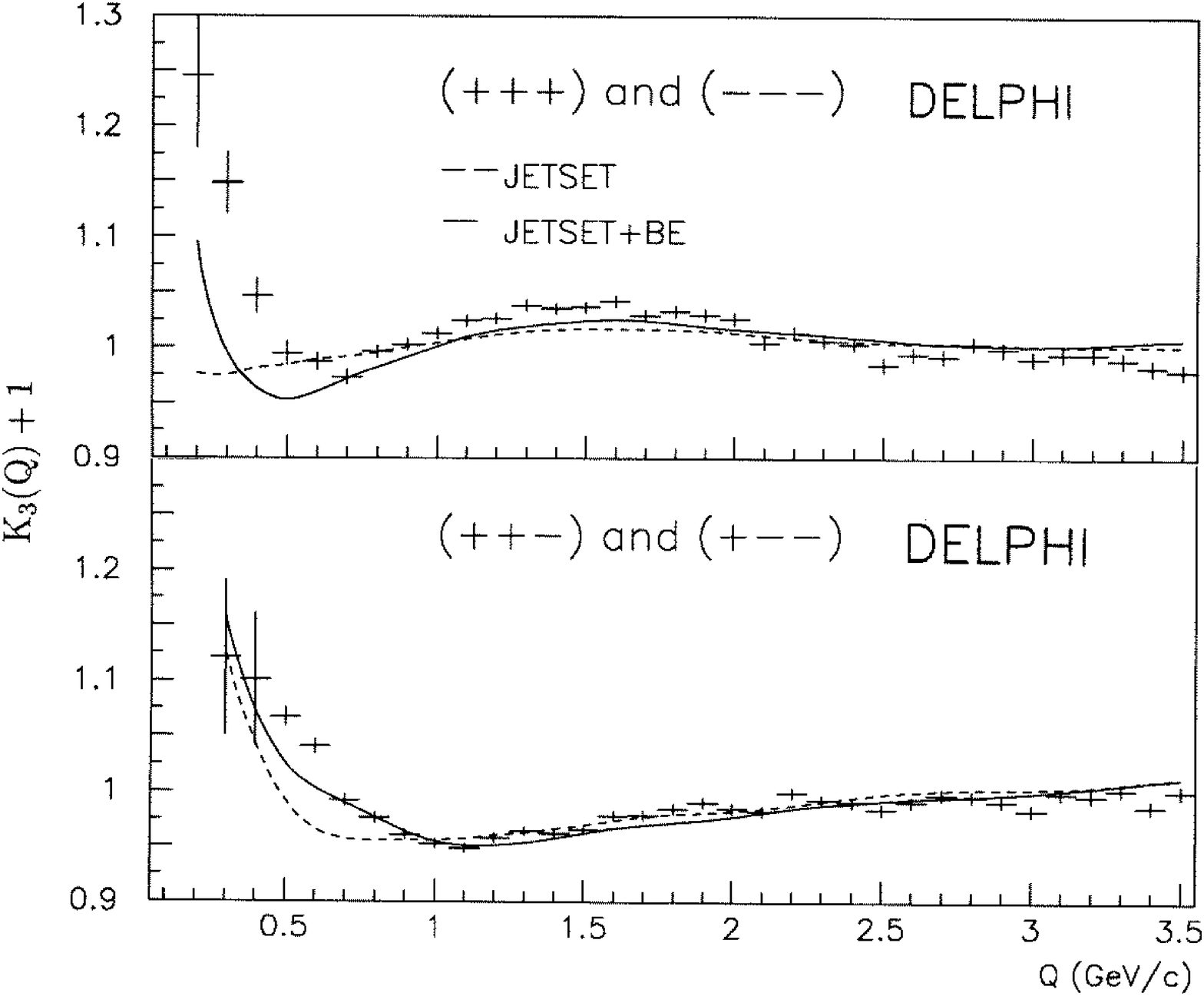,width=7.5cm}}
\vs 2mm
{\footnotesize\baselineskip=10pt\ni
Figure 6. Genuine three-particle correlation in hadronic Z decay, compared
to JETSET with and without LUBOEI.\protect\cite{25}
\par}
\end{figure}

\vs 2mm\ni
6. {\em Density (or multiplicity) dependence}:
A linear increase of the size of the pion-emission region with increasing
particle density, combined with decrease of the correlation-strength
parameter $\la$ is well known from heavy-ion and higher-energy ISR
and collider results (e.g. \cite{27} and refs. therein).
At least the decrease of $\la$ can be understood from the overlap of an
increasing number of independent mechanisms (e.g. strings or clusters).
OPAL \cite{28} has shown that a similar dependence is also present in Z
decay (Fig.~7). It can at least in part be explained from the presence of 
two- and three-jet events.

\begin{figure}
\begin{minipage}[b]{7cm}
\begin{center}
\epsfig{bbllx=77pt,bblly=150pt,bburx=464pt,bbury=722pt,clip=,file=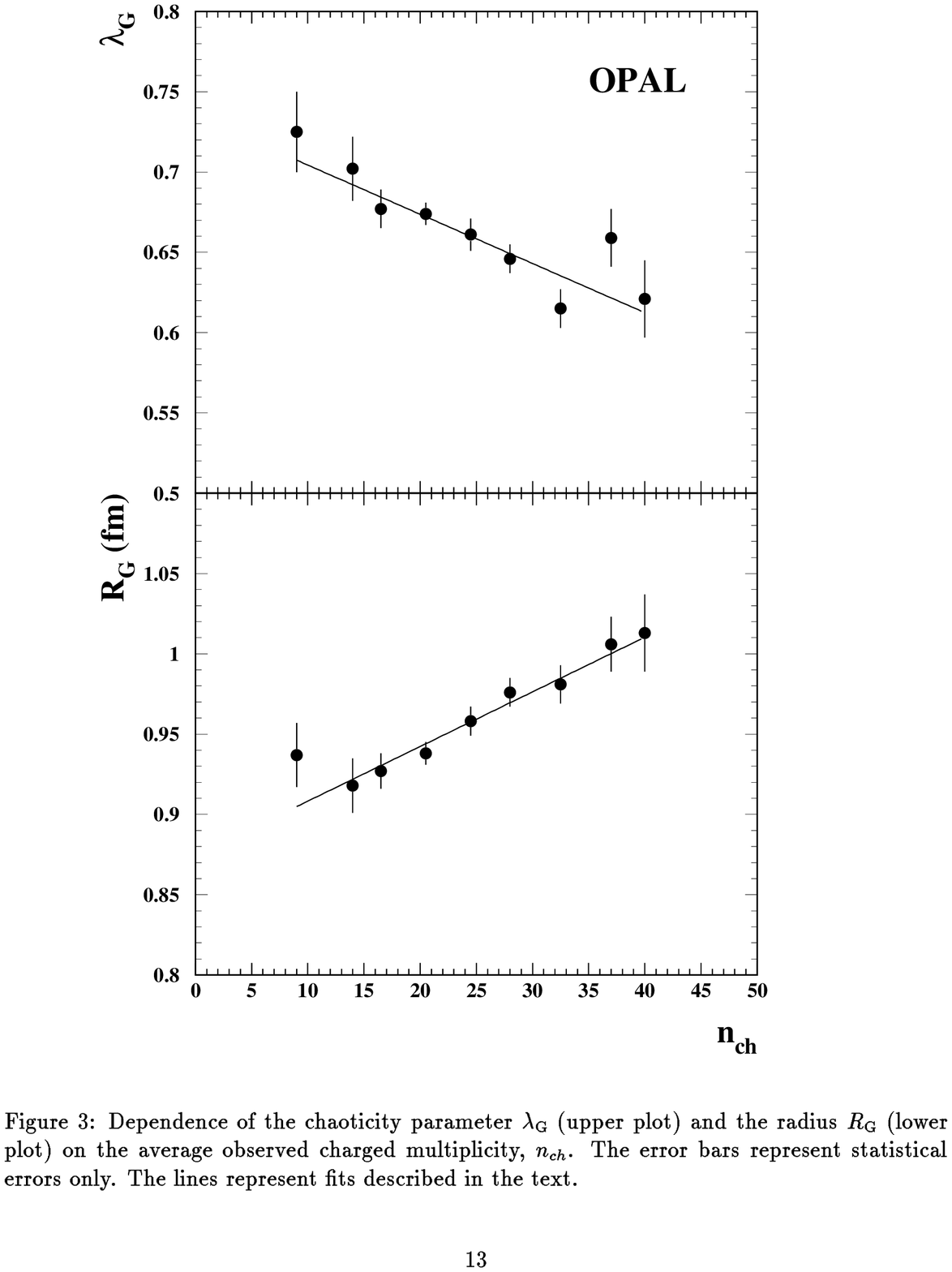,width=6.5cm}
\end{center}
{\footnotesize\baselineskip=10pt\ni
Figure 7. 
Dependence of $\la$ and $r$ on the charged-particle
multiplicity $n$ for e$^+$e$^-$ collisions at the Z mass.\protect\cite{28}\par}
\end{minipage}
\hs 1cm
\begin{minipage}[b]{7cm}
\cent{\epsfig{file=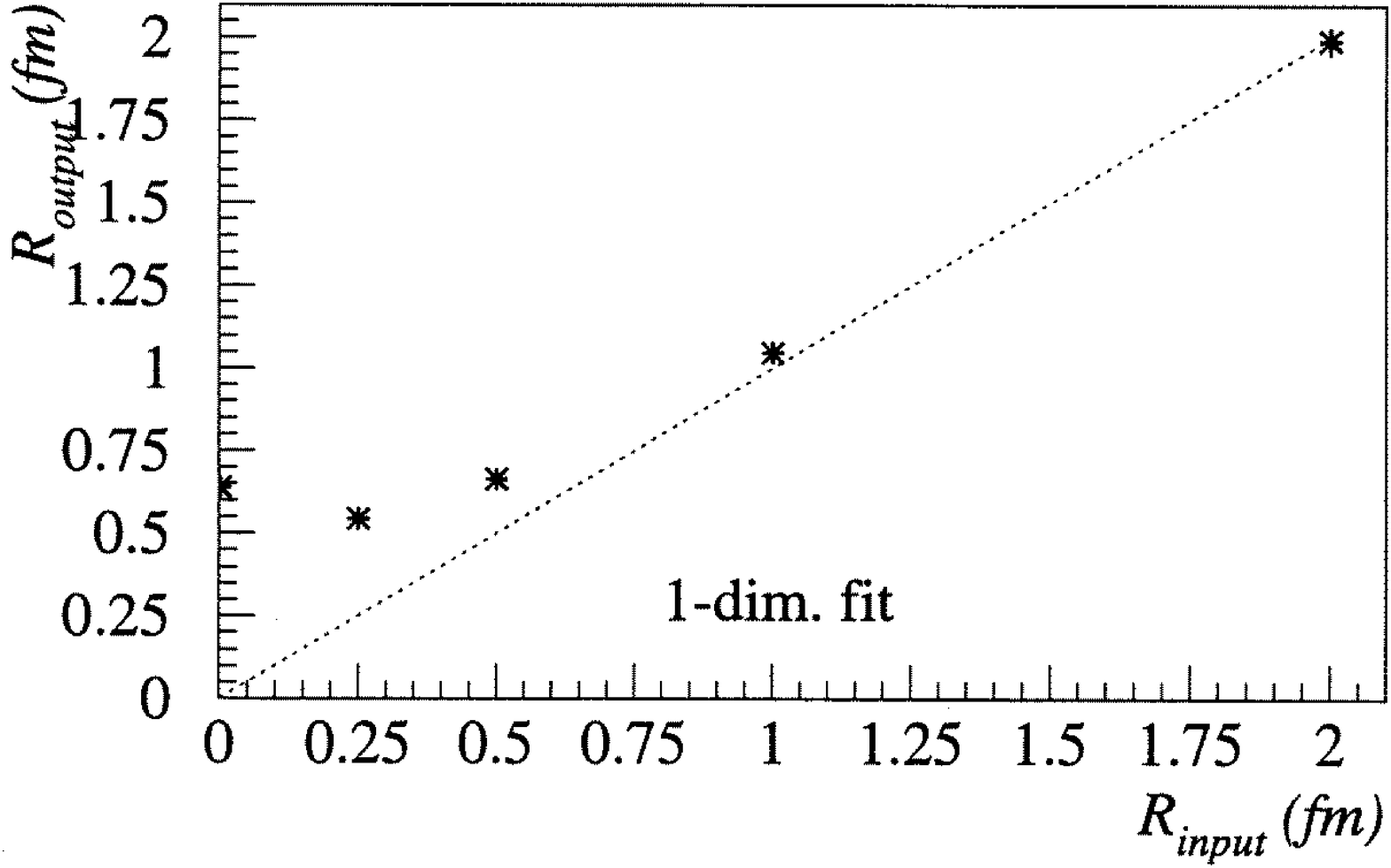,width=6.5cm}}
\cent{\epsfig{file=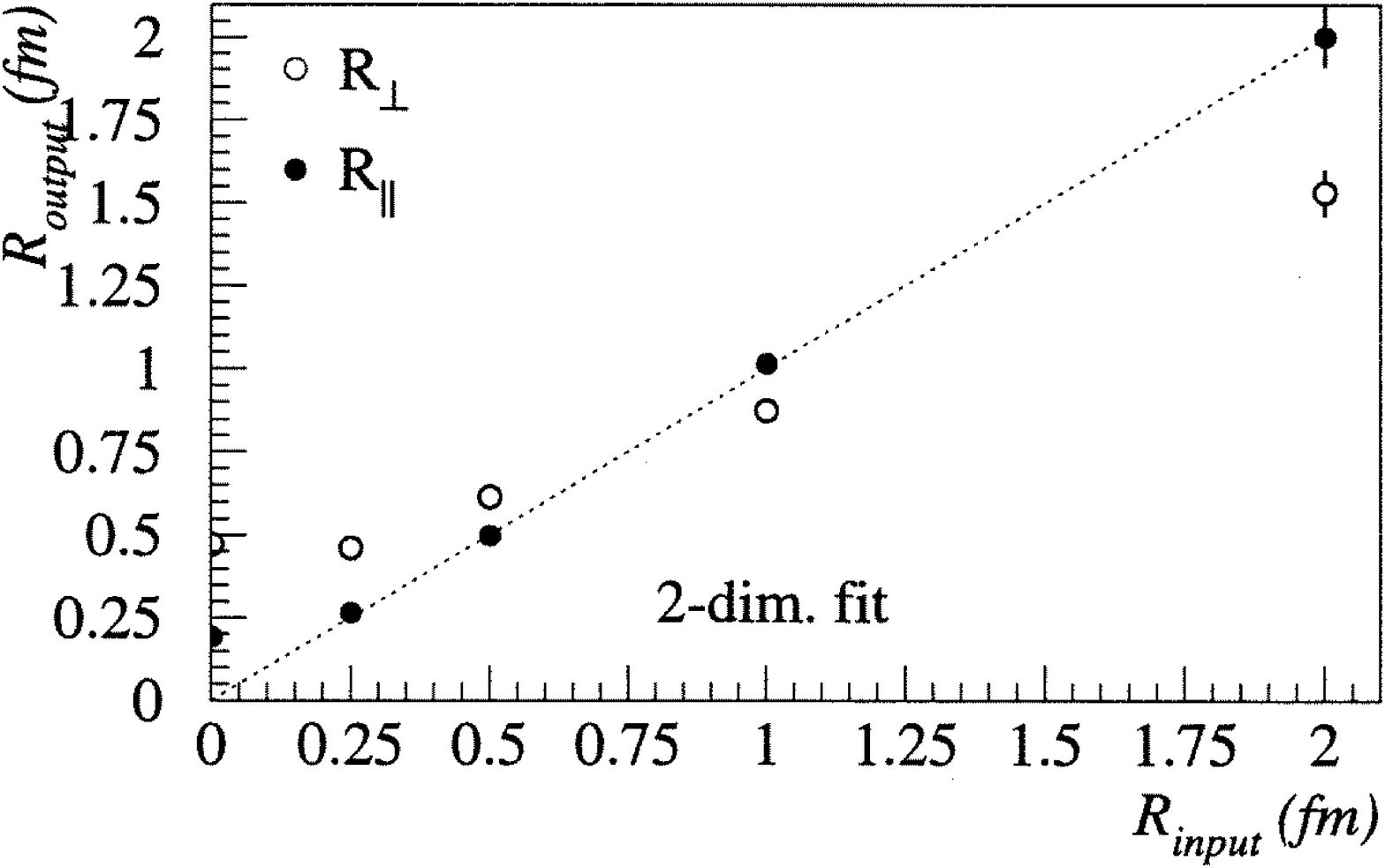,width=6.5cm}}
\vs 5mm
{\footnotesize\baselineskip=10pt\ni
Figure 8. LUBOEI output radius parameter $\rr_\output$ value as a function
of the input parameter $\rr_\input$.\protect\cite{31} 
\par}
\end{minipage}
\end{figure}

\subsection{Three types of Monte-Carlo implementation}
\vs 2mm\ni
1. {\em Reshuffling}:
The MC code LUBOEI \cite{29} in JETSET treats BE correlations as a final
state interaction (!) and actually changes particle momenta according to a 
spherically symmetric Gaussian (or, alternatively, exponential) correlator.
The advantages are that it is a fast and unit-weight (i.e. efficient)
generator.
The bad news are that it is imposed a-posteriori (without any physical basis),
is even unphysical (since it changes the momenta), is not self-consistent
(since it introduces an artificial length scale \cite{30,31}), see Fig.~8, 
is spherically symmetric, does not treat higher-order correlations properly,
etc. The worse news is that it is used by everybody to correct for detector
effects and that there is no perfectly tested alternative, at the moment.

\vs 2mm\ni
2. {\em Global reweighting}:
Another, theoretically better justified approach is to attach to each
pre-generated event a BE weight depending on its momentum configuration, 
but leaving this momentum configuration untouched.
Based on the use of Wigner functions \cite{32} rather than amplitudes,
a weight factor can be derived of the form \cite{33} 
\be
W(p_1,\dots p_n)= \sum_{\{P_n\}} \prod^n_{i=1} K_2 (Q_{iP_n(i)})
\ee
where $n$ is the number of
identical particles, $K_2(=R_2-1)$ is the two-particle correlator 
and $P_n(i)$ is the particle which occupies the position $i$ in the
permutation $P_n$ of the $n$ particles.
Applications of the global weighting \cite{34,35,36,37,38} are
essentially all variations on this theme, with varying model assumptions
on the exact form of $K_2$.
In general, $K_2(Q)$ is still assumed to be spherical in $Q$, even 
though a generalisation would be simple to implement. Higher-order
correlations are, in principle, included, but either assume \cite{36} a
quantum optical model, already shown to be wrong,\cite{24,39} or
factorization in terms of Eq.~(5) not allowing for phases between the terms.

More seriously, as in \cite{29} the weight is imposed a posteriori on a MC 
event pre-fabricated according to a given model, so not as a part of this 
model, itself.

Problems arise from the fact that the number of permutations is $n$!
so that simplifications have to be introduced.\cite{38} Wild fluctuations
of event weights can occur, so that cuts on event weight are necessary.
The weight may even change the parton distributions, while BE correlations
only work on the pion level. Retuning is of course necessary, but this can, in
practice be achieved by just retuning the multiplicity distribution.\cite{38}
 
\vs 2mm\ni
3. {\em Symmetrizing}:
Bose-Einstein correlations have been introduced
into string models, directly.\cite{40,41} In these models, an ordering in 
space-time exists for the hadron momenta within a string.
Bosons close in phase space are nearby in space-time and 
the length scale measured by Bose-Einstein correlations is not
the full length of the string, but the distance in boson-production
points for which the momentum distributions still overlap.

The (non-normalized) probability $\rd\G_n$ to
produce an $n$-particle state $\{ p_j \}$, $j=1,\dots n$ of distinguishable
particles is
\be
\rd \G_n=[\Pi^n_{j=1} N \rd p_j \d(p^2_j-m^2_j)] \d(\S p_j-P)\exp (-bA_n)\ ,
\ee
where the exponential factor can be interpreted as the square of a matrix 
element $M_n = \exp (i\xi A_n), \rR\re (\xi)=\k, \rI\rrm (\xi)=b/2$, and the 
remaining terms describe phase space, with $P$ being the total 
energy-momentum of the state. $N$ is related to the mean multiplicity and $b$ 
is a decay constant related to the correlation length in rapidity. $A_n$
corresponds to the total space-time area covered by the color field, 
or to an equivalent area in energy-momentum space divided
by the square of the string tension $\k=1$ GeV/fm.\cite{41}

\begin{figure}[t]
\begin{center}
\epsfig{file=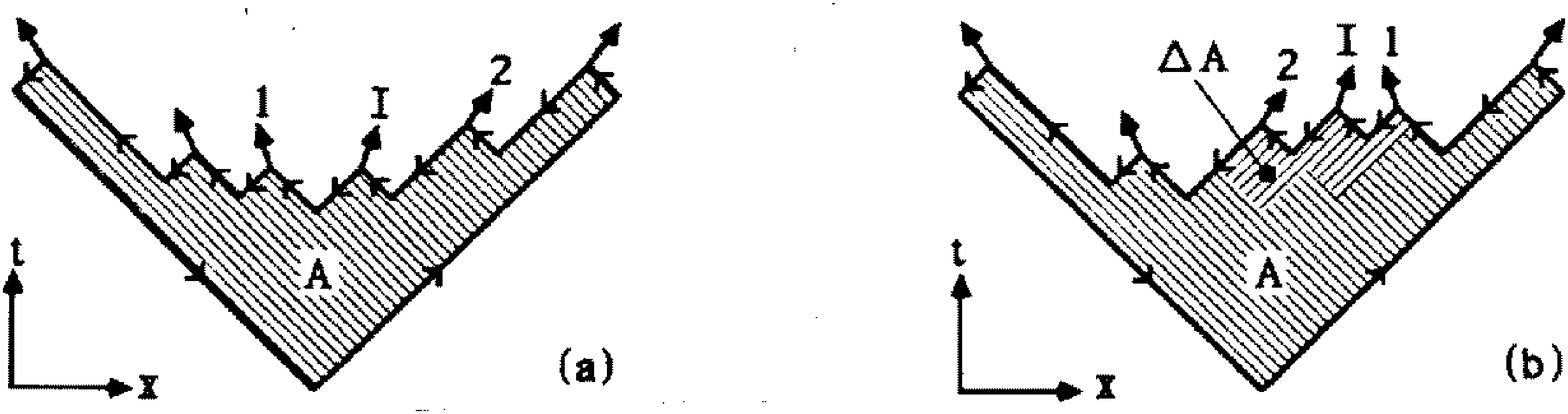,width=12cm}
\end{center}
\vs-3.5cm
{\footnotesize\baselineskip=120pt\ni
Figure 9. Space-time diagram for two ways to produce two identical
bosons in the color-string picture \protect\cite{40}.\par}
\end{figure}

The production of two identical bosons (1,2) is governed by the 
symmetric matrix element
\be
M=\frac{1}{\sqrt 2} (M_{12}+M_{21})= \frac{1}{\sqrt 2} [\exp (i\xi A_{12})+
\exp (i\xi A_{21})]\ . 
\ee
There is an area difference and, consequently, a phase difference between 
$M_{12}$ and $M_{21}$ of $\D A=|A_{12}-A_{21}|$,
where the indices 1, 2 particles 1, 2, respectively (see Fig.~9).

Using this matrix element, one obtains
\be
R_\BE\approx 1+\lan \cos(\k\D A)/\cosh (b\D A/2)\ran\ \ ,
\label{13-13b3} 
\ee
where the average runs over all intermediate systems I. In the limit 
$Q^2=0$ follows, $\D A=0$ and $R_\BE= 2$,
in agreement with the results from the conventional interpretation for
completely incoherent sources. However, for $Q^2\not= 0$ follows an additional
dependence on the momentum $p_\rI$ of the system I produced between
the two bosons. 

The model can account well for most features of the e$^+$e$^-$ data, 
including the non-spherical shape of the
BE effect. More recently, the symmetrization has been generalized to more 
than 2 identical particles.\cite{45} This approach deserves strong support. 
A more detailed account will be given in the next talk.\cite{18}

\section{Conclusions}

With respect to color reconnection, my view is that VNI is out,
that no effect has been observed in WW decay with the variables used so far,
but that more discriminative methods , as those applied in correlation and
fluctuation analysis, have to be used.

With respect to BE correlations, I conclude, they may form a problem,
but also can be used to study the very space-time development of the WW 
overlap. Since this first needs a detailed study of the space-time 
development of a single high-energy $\rq_1\bar\rq_2$ system, I suggest 
(in parallel to continued direct WW analysis) a four-step program for an 
analysis of the final data to come:

1. Look at the Z in much more detail. In fact, a lot more information is
available or becoming available than used by most of the model builders.
E.g., the elongated, non-Gaussian shape of the correlation function excludes
the present version of all models, except those of \cite{18,40,41,45}. 
The shape of the emission function for a single $\rq_1\bar\rq_2$ system 
in space-time determines the actual WW overlap. This shape is known for 
hh and heavy-ion collisions and should be urgently measured at the Z. 
Higher-order correlations, a density dependence and a transverse-mass 
dependence are observed and can be expected to discriminate between models.

2. Tune the models passing these tests on the Z, with and without b-quark 
contribution.

3. Check them on a single W.

4. Only then apply them to WW decay.
 
One important last point: color reconnection and Bose-Einstein effects can 
(partially) cancel, as e.g. in multiplicity. So, in fully hadronic WW decay, 
their effects have definitely to be studied simultaneously, in the data,
as well as in the models!

\end{document}
